# A Full Duplex Transceiver with Reduced Hardware Complexity

Mustafa Emara*, Patrick Rosson†, Kilian Roth*, David Dassonville†
* Next Generation and Standards, Intel Deutschland GmbH, Neubiberg, Germany
Email: {mustafa.emara, kilian.roth}@intel.com
† CEA, LETI, MINATEC Campus, F-38054 Grenoble, France
Email: {patrick.rosson, david.dassonville}@cea.fr

*Abstract*—For future wireless communication systems, full duplex is seen as a possible solution to the ever present spectrum shortage. The key aspect to enable In-Band Full Duplex (IBFD) is sufficient cancellation of the unavoidable Self-Interference (SI). In this work we evaluate the performance of a low complexity IBFD transceiver, including the required analog and digital interference cancellation techniques. The Radio Frequency Self-Interference Canceler (RFSIC) is based on the isolation of a circulator in combination with a vector modulator regenerating the interference signal, to destructively combine it with the received signal. On the digital side, a Digital Self-Interference Cancellation (DSIC) algorithm based on non-linear adaptive filtering is used. With the simplified analog front-end of a Software Defined Radio (SDR) platform, SI cancellation of 90 dB is achieved with the presence of a received signal.

## I. INTRODUCTION

Increasing the spectral efficiency is crucial for cellular systems, especially in the context of spectrum scarcity below 6 GHz. IBFD transceivers represent a mean to achieve this goal, since the same time-frequency resources are used for both transmission and reception [1]. Moreover, IBFD provides a theoretical spectral efficiency gain of up to two, for a point-to-point link compared to a Frequency Division Duplex (FDD) system, by increasing the available bandwidth if certain conditions are met [2]. Compared to a Time Division Duplex (TDD) system, the number of available time-slots for Up Link (UL) and Down Link (DL) is increased.

It has been shown in [3] that significant gains are obtained at the system level, even with limited interference cancellation capabilities. Even for a system that only achieves 65 dB of Self-Interference Cancellation (SIC), it is possible to improve the average downlink and uplink throughput by 21 % and 4.9 % respectively. This performance improvement increases to 69 % and 81 % for 85 dB of cancellation.

The main challenge of IBFD is the cancellation of the SI resulting from the transmit signal and leaking into the receiver front-end. Since the difference between the SI and the received signal could easily exceed 90 dB, several stages are required to reduce the SI power.

Firstly, different antenna concepts can reduce the SI power. High isolation can be provided either by using a circulator with well adapted antennas or dual-polarized antennas for transmission and reception [4]. Utilizing the former solution assumes channel reciprocity in point to point scenarios, whereas the latter addresses distinctive users with different channels [5].

Secondly, to avoid saturation of the Low Noise Amplifier (LNA), it is important to add a RFSIC that attenuates the SI at the receiver's front-end. This is usually achieved by a combination of passive and/or active techniques [6][7]. The latter aims at regenerating the SI signal and then destructively combining it with the actual received signal. On the other hand, the passive techniques are generally based on filtering (narrow or broadband) of the transmitted signal before subtracting it from the remaining received signal.

After the RFSIC, the residual signal is mixed to the analog baseband and then digitalized. It is important that the dynamic range of the Analog to Digital Converter (ADC) covers the superposition of the SI and the Signal of Interest (SoI). Moreover, the RFSIC should not add non-coherent noise that cannot be removed throughout the SIC process. Moving to the DSIC, its objective is to cancel the remaining SI down to the system's noise floor. Therefore, it is necessary to properly cancel the remaining linear as well as the nonlinear components of the SI signal.

It was shown in [8] that a 110 dB SIC level was reached over an 80 MHz bandwidth at 2.4 GHz. Nevertheless, this solution includes a complex and expensive 16-tap Radio Frequency (RF) filter. The setup in [9] considered a relatively small bandwidth of 625 kHz, but provided up to 100 dB of self interference cancellation, including 41 dB of antenna isolation by using separate Tx and Rx antennas.

The objective of this work is to reach an acceptable SIC level, with a less complex solution. For this setup, a classical SDR radio is used with only 12 bits resolution for both the Digital to Analog Converter (DAC) and the ADC. In contrast to previous work, the investigation was extended to cover the presence of a received SoI.

The paper is organized as follows. In Section II, the design of the implemented RFSIC along with the algorithm of the DSIC are presented. Additionally, the system setup and the evaluation results are discussed in Section III. Finally, conclusion of the work is summarized in Section IV.

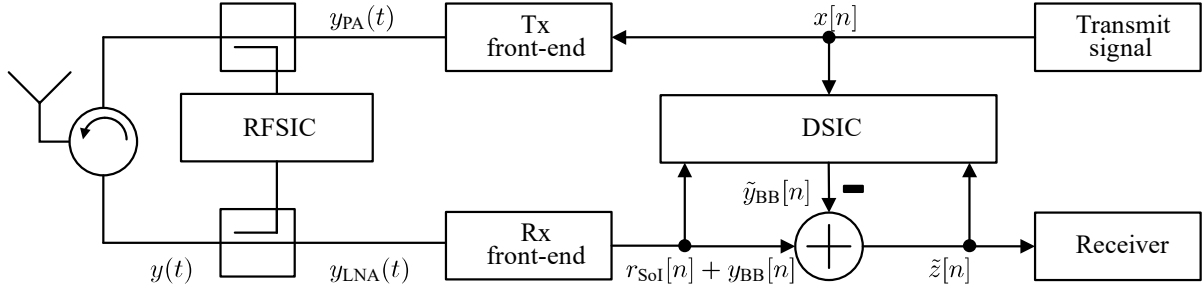
Fig. 1. Full duplex transceiver system diagram.

## II. FULL DUPLEX TRANSCEIVER DESIGN

### A. System Model

The system model investigated throughout this work is shown in Figure 1. The generated baseband signal denoted by $x[n]$ first propagates through the transmitter's front-end, resulting in the analog signal at the Power Amplifier (PA) output $y_{\text{PA}}(t)$. Afterwards, the signal passes via a directional coupler to the circulator for transmission and the coupled signal goes towards the RFSIC function. The received signal denoted by $y(t)$, includes the SoI along with the SI signal. After the RFSIC process, the signal $y_{\text{LNA}}(t)$ denotes the input to the LNA. This signal is then processed by the receiver, resulting in the digital baseband signal $r_{\text{SoI}}[n] + y_{\text{BB}}[n]$ which incorporates the residual linear and nonlinear SI $y_{\text{BB}}[n]$, and the received SoI $r_{\text{SoI}}[n]$. The DSIC block is then utilized to compute an estimate of the SI signal denoted by $\tilde{y}_{\text{BB}}[n]$ which is subtracted from the actual SI signal resulting in the residual signal $\tilde{z}[n]$ after the digital cancellation stage.

### B. Radio Frequency Self-Interference Cancellation

The RFSIC spans three main requirements; considerable SI reduction, low noise addition and stable signal tracking. The first requirement aims at avoiding the saturation of the receiver's front-end, especially the ADC [10]. Accordingly, the receiver gain needs to be adapted to the smaller received signal's power. In this paper, the usage of the circulator was employed as a solution to this aspect. The second important requirement addresses the added noise by the RFSIC. If the RFSIC introduces additive non-coherent noise, its digital subtraction will be impossible, due to the random nature of the noise. The last requirement targets the oscillations existing in the hardware circuit. The last requirement targets a stable SI signal tracking capability. The RFSIC has a capacity to follow the SI evolution in time, however, it should change slowly to allow the digital canceler to converge. Moreover, any power supply harmonics that induces fast oscillations at the RF canceler must be perfectly filtered. Globally, the RFSIC must reduce the SI in order to use the receiver gain in the correct setting and not degrade the receiver's noise figure.

Without the RFSIC, the SI received at the LNA's input can be modeled as

$$y_{\text{LNA}}(t) = \sum_{\alpha} h_\alpha y_{\text{PA}}(t - \tau_\alpha) + \sum_{\beta} h_\beta y_{\text{LNA}}(t - \tau_\beta). \quad (1)$$

The terms $h_\alpha y_{\text{PA}}(t - \tau_\alpha)$ correspond to the multi-path components coming from the PA and going towards the LNA. It consists of circulator leakage, antenna return loss and over the air scattered wave close to the antenna [11]. The second term $h_\beta y_{\text{LNA}}(t - \tau_\beta)$ corresponds to the two way traveling wave due to the return loss of the receiver and the transmitter. It is observed that $h_\beta y_{\text{LNA}}(t - \tau_\beta) < h_\alpha y_{\text{PA}}(t - \tau_\alpha)$ due to the adopted transceiver's architecture. However, the second term is not negligible if 100 dB of cancellation is considered, as it provides room for further enhancement when modeled.

In order to reduce the SI, the RF canceler through this paper is proposed to be added between the output of the PA and the input of the LNA. Additionally, a directional coupler is inserted at the output of the PA to capture the transmitted signal. This signal is then delayed with a fixed delay and afterwards attenuated and rotated by a vector modulator. The vector modulator's output is then injected via another directive coupler before the LNA. Due to the limited directivity of the couplers or its return loss, other undesired signals are injected in the RF canceler. It is important to note that directive couplers induce loss on the direct path which reduce the transmitted power on the Tx side, and degrade the noise figure on the Rx side. To combat this drawback, high directivity couplers are advised. To avoid the usage of an active amplifier in the RFSIC path, the overall minimum loss on the RFSIC must be lower than the antenna system isolation.

With an active amplifier in the RFSIC working in the linear zone, the SI signal received at the LNA can be modeled as

$$y_{\text{LNA}}(t) = \sum_{\alpha'} h_{\alpha'} y_{\text{PA}}(t - \tau_{\alpha'}) + \sum_{\beta'} h_{\beta'} y_{\text{LNA}}(t - \tau_{\beta'}) + b(t), \quad (2)$$

where $b(t)$ models the noise generated when the RFSIC is added. The vector modulator is controlled in order to reduce the overall received power. It can be observed that with one fixed delay and one variable complex path, it is impossible to cancel all the leakage paths present in the initial configuration. Nevertheless, when the delay is tuned accordingly and the vector modulator is correctly set, the stronger leakage paths can be mitigated. It is worth mentioning that if the RFSIC works in the nonlinear zone, the previous model needs to be upgraded, taking into account the added nonlinear effect. Finally, the analog signal $y_{\text{LNA}}(t)$ is frequency down converted and then A/D converted. Through the coming part, the DSIC's design will be presented.

## C. Digital Self-Interference Cancellation

The main objective of the digital cancellation is to cancel the residual SI after the analog cancellation. This includes the nonlinear components, in particular the remaining nonlinearities added by the RFSIC and the circulator [8].

Throughout this work, a nonlinear adaptive digital cancellation solution based on the work in [12], utilizing the transversal recursive least squares is adopted. A pre-adaptation orthogonalization based on the Cholesky decomposition is carried out to further enhance the adaptation process.

A proper modeling of the SI signal affects the system design and performance. Inspired by [8] and [13], the overall SI baseband signal can be modeled using a parallel Hammerstein model

$$y_{\text{BB}}[n] = \sum_{m=0}^{M-1} \sum_{p=1}^{P} h_p[m] \phi_p(x[n-m])$$
$$\phi_p(x[n]) = x[n]|x[n]|^{p-1}, \quad (3)$$

where $M$ depicts the memory depth of the model and $P$ is the nonlinearity order. The symbol $h_p[m]$ represent the $p$th order channel coefficients of the effective SI channel and $\phi_p(x[n])$ denotes the nonlinear basis function of the baseband signal $x[n]$.

The signal after the DSIC depends on the SI channel coefficients $\tilde{h}_p[m]$, the received signal of interest $r_{\text{SoI}}[n]$ and the existent Additive White Gaussian Noise (AWGN) in the receiver $\eta[n]$ as follows

$$\tilde{z}[n] = r_{\text{SoI}}[n] + y_{BB}[n] + \eta[n] - \tilde{y}_{BB}[n]$$
$$= r_{\text{SoI}}[n] + \eta[n] +$$
$$\underbrace{y_{BB}[n] - \sum_{m=0}^{M-1} \sum_{p=1}^{P} \tilde{h}_p[m] \phi_p(x[n-m])}_{\text{Residual SI}}. \quad (4)$$

The main target is to provide an accurate, fast and low-complex estimation of the SI channel coefficients in order to regenerate the SI signal.

Since $P$ nonlinear basis functions are generated for every incoming sample, the basis functions across different nonlinearity orders are highly correlated and have different variance. Thus, a slow convergence and a degraded cancellation performance can be observed while estimating the SI effective channel coefficients. Consequently, an orthogonalization of the basis functions before the coefficients estimation is required [13]. The covariance matrix of the basis functions across sufficiently large number of samples can be computed as

$$\boldsymbol{\Upsilon} = \mathbb{E}[\boldsymbol{\phi}[n] \boldsymbol{\phi}^{\text{H}}[n]], \quad (5)$$

where $\mathbb{E}[\cdot]$ is the expectation operation. The vector $\boldsymbol{\phi}[n] = [\phi_1[n] \; \phi_2[n] \ldots \phi_P[n]]^{\text{T}}$ represents the instantaneous basis functions of the $n$-th sample. A transformation of the basis function is carried out via a whitening transformation matrix $\boldsymbol{T}$ based on the Cholesky decomposition

$$\boldsymbol{\Upsilon} = \boldsymbol{L}\boldsymbol{L}^{\text{H}},$$
$$\boldsymbol{T} = \boldsymbol{L}^{-1}, \quad (6)$$

where $\boldsymbol{L}$ is a lower triangular matrix with positive diagonal entries. The orthogonalized basis functions $\tilde{\boldsymbol{\phi}}[n]$ are computed as

$$\tilde{\boldsymbol{\phi}}[n] = \boldsymbol{T}\boldsymbol{\phi}[n]. \quad (7)$$

In order to simplify the notation of the signal model in (4), it can be reformulated such that the data vector for the previous $M$ samples are included as

$$\boldsymbol{u}[n] = [\tilde{\boldsymbol{\phi}}^{\text{T}}[n] \; \tilde{\boldsymbol{\phi}}^{\text{T}}[n-1] \ldots \tilde{\boldsymbol{\phi}}^{\text{T}}[n-M+1]]^{\text{T}} \in \mathbb{C}^{MP \times 1}, \quad (8)$$

where $\boldsymbol{u}[n]$ is the input complex data vector. Applying the same notation to the estimated SI channel coefficients $\tilde{\boldsymbol{h}}[n]$ can get

$$\tilde{\boldsymbol{h}}[n] = [\tilde{h}_1[n] \; \tilde{h}_2[n] \ldots \tilde{h}_P[n] \ldots \tilde{h}_P[n-M+1]]^{\text{T}} \in \mathbb{C}^{MP \times 1}, \quad (9)$$

where $\tilde{\boldsymbol{h}}[n]$ are the SI channel coefficient to be estimated. From plugging (8) and (9) into (4), the residual signal after the DSIC can be reformulated to

$$\tilde{z}[n] = r_{SoI}[n] + \eta[n] + y_{BB}[n] - \tilde{\boldsymbol{h}}^{\text{H}}[n]\boldsymbol{u}[n]. \quad (10)$$

The proposed algorithm in this work is based on the Recursive Least Squares (RLS) algorithm combined with complexity reduction techniques. Inspired by the work in [14], the summary of the proposed exponentially weighted RLS algorithm is presented in Table I.

In the initialization step residual vector $\boldsymbol{r}[n]$ is set to the covariance vector $\boldsymbol{\beta}_o[n]$. The correlation matrix $\boldsymbol{R}[n]$ is set to an equalization matrix $\boldsymbol{\Pi} = \alpha \boldsymbol{I}_{MP}$, where $\boldsymbol{I}_{MP}$ is an identity matrix of dimension $MP \times MP$. The parameter $\alpha$ is chosen based on the Signal-to-Noise-Ratio (SNR) as $0 < \alpha < 1$ [15]. The parameter $\lambda$ is the forgetting factor and is chosen as $0 << \lambda \leq 1$.

The first step represents the update of the correlation matrix for each incoming sample. Originally, the update should consider all the coming input data vector $\boldsymbol{u}[n]$. Nevertheless, following the stationarity assumption of the input data, only the first $p$ components of the data vector are sufficient to reconstruct the complete correlation matrix. Those $p$ components are fully captured in $\tilde{\boldsymbol{\phi}}[n]$. The notation $\boldsymbol{R}^{(1:p)}[n]$ stands for the first $p$ rows of the correlation matrix. Thus, by exploiting the transversal structure of the input data vector, a reduction in the computational complexity is achieved.

An efficient solution should be utilized to solve step 4, which constitutes the complexity bottleneck of the algorithm.

TABLE I
EXPONENTIALLY RECURSIVE LEAST MEAN SQUARES ALGORITHM

| Step | Computation | real × | real + |
|---|---|---|---|
| | **Initialization**: $\tilde{\boldsymbol{h}}[0] = \boldsymbol{0}, \tilde{z}[0] = 0$ $\boldsymbol{r}[0] = \boldsymbol{0}, \boldsymbol{R}[0] = \boldsymbol{\Pi}$ | | |
| | while transmitting $(n \geq 1)$ | - | - |
| 1 | $\boldsymbol{R}^{(1:p)}[n] = \lambda \boldsymbol{R}^{(1:p)}[n-1] + \tilde{\boldsymbol{\phi}}[n]\boldsymbol{u}^{\mathrm{H}}[n]$ | $6MP^2$ | $4MP^2$ |
| 2 | $\tilde{z}[n] = y_{\mathrm{BB}}[n] - \tilde{\boldsymbol{h}}^{\mathrm{H}}[n-1]\boldsymbol{u}[n]$ | $4MP$ | $2(MP+1)$ |
| 3 | $\boldsymbol{\beta}_o[n] = \lambda \boldsymbol{r}[n-1] + \tilde{z}^*[n]\boldsymbol{u}[n]$ | $6MP$ | $4MP$ |
| 4 | $\boldsymbol{R}[n]\Delta\tilde{\boldsymbol{h}}[n] = \boldsymbol{\beta}_o[n] \Rightarrow \Delta\tilde{\boldsymbol{h}}[n], \boldsymbol{r}[n]$ | $P_M$ | $P_A$ |
| 5 | $\tilde{\boldsymbol{h}}[n] = \tilde{\boldsymbol{h}}[n-1] + \Delta\tilde{\boldsymbol{h}}[n]$ | - | $2MP$ |
| | Total: × : $6MP^2 + 10(MP) + P_M$ Total + : $4MP^2 + 8MP + 2 + P_A$ | - | - |

This step results in computing the coefficients update step $\Delta\tilde{\boldsymbol{h}}[n]$ along with the residual vector $\boldsymbol{r}[n]$. The symbols $P_M$ and $P_A$ stand for the complexity of real multiplications and additions of step 4. The focus has been directed throughout this work towards the Dichotomous Coordinate Descent (DCD) algorithm due to its low complexity advantage [16]. It has been shown that the maximum number of additions $P_A$ required is upper bounded by $N(2N_u + M_b - 1) + N_u$ and the number of multiplications $P_M$ is zero.

## III. SYSTEM EVALUATION

### A. Platform Overview

The test bench presented in this section allows the performance evaluation of the RFSIC and the DSIC on real signal with the presence of an external received signal. A block diagram illustrating the designed test bench is shown in Figure 2. A PC running Matlab is connected to a digital board that supports simultaneous Tx/Rx data flows. The digital board is connected to a SDR ARRadio board. This RF board includes an AD9361 SDR chip from Analog Devices as the RF transceiver. It supports the converters (ADC/DAC), the frequency conversion (up/down) as well as a driver amplifier and a LNA. The output RF signal is then connected to an additional PA and the RFSIC board. As presented earlier, we consider the circulator solution, where the antenna has only one input/output port. The antenna is replaced by a static passive emulator circuit which emulates the reflection seen on an antenna port.

At the system level, we consider that the circulator belongs to the antenna sub-system. We recall that in IBFD system consideration, two options are possible depending on the scenario; either one single-port antenna with a circulator, or a two port antenna addressing two antenna patterns (one for Tx, one for Rx). Two separate antennas falls into the same scenario as a two port antenna. Throughout the following subsection, the specific elements of the test setup are presented.

*1) Antenna emulation with circulator:* The usage of an antenna emulation circuit represents an alternative to a real antenna. Such a choice was motivated by the avoidance of the uncertainty caused by the time-varying environment, reflections and changing channel propagation, which would increase the complexity of the system debugging. The antenna emulation circuit is an RF passive circuit which is constant over time and has a known return loss. This solution is a first step to investigate the different interference cancellation stages. The time varying reflections with a real antenna will be studied in a second step.

Figure 3 presents the RF leakage due to the combination of the circulator and the antenna emulation circuit. The circulator leakage with antenna emulation is approximately -20 dB at the center frequency of 900 MHz and varies about 1 dB across a bandwidth of 20 MHz. Additional time varying RF leakage would originate from other interference such as power supply coupling, active circuit coupling and the internal components of the SDR chip.

*2) RFSIC board:* The RF canceler board connects the RF transceiver board on one side and the circulator on the other side. Inside the RFSIC, a sample of the Tx signal is vector-modulated, delayed and amplified, before being added to the Rx signal. The vector-modulator is controlled by two analog voltages of the digital board.

*3) Transceiver board:* The RF transceiver board is based on the ARRadio board. It embeds an AD9361 SDR chip [17]. A PA is added to increase the transmitted power to 20 dBm. Due to the wide-band capability of the SDR Integrated Circuit (IC), external band filters are necessary to prevent aliasing. As the AD9361 has a good noise figure, no external LNA is required.

*4) Digital Board:* The digital board controls the configuration of the AD9361; especially the Phase Lock Loop (PLL) and the transmitter and receiver gain. It is able to send the IQ data to the transceiver board and synchronously receive the IQ data of the received signal from the transceiver board. This board has been designed to support simultaneously Filter Bank Multi-Carrier (FBMC) transmission and reception and has the additional capacity to manage IBFD. The digital board is also able to generate two DC voltages to control the RFSIC.

A Graphic User Interface (GUI) running on the PC allows to activate different SI cancelers for the IBFD modes (i.e. time-frequency synchronized samples generation and acquisition). The data has up to 12 bits resolution for IQ data. In this paper, we will focus on tests in the 900 MHz frequency band. The RF bandwidth is set to 56 MHz and the sample rate is set to 61.44 MHz. In this configuration, the chip works in FDD mode but the Tx Local Oscillator (LO) and the Rx LO are set to the same frequency. It is also assumed that the transmission and reception are synchronized in time.

### B. Results

*1) RFSIC performance:* This first showcase presents the RFSIC performance. Figure 4 shows the Power Spectrum Density (PSD) measured by a FSW signal analyzer of the transmitted signal over multiple stages against the relative frequency in MHz. It is observed the RFSIC's effect and that with the proposed solution, a value of 36 dB of cancellation was realized. Two other signals are measured (noise floor of the RFSIC in black, noise floor of the signal analyzer in yellow) to indicate the cancellation limits imposed by the employed hardware components.

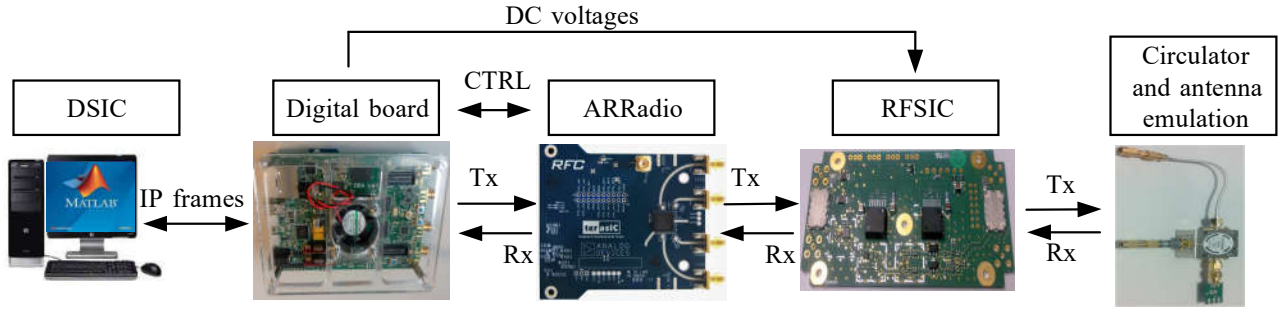

Fig. 2. System setup incorporating passive cancellation, RFSIC and DSIC.

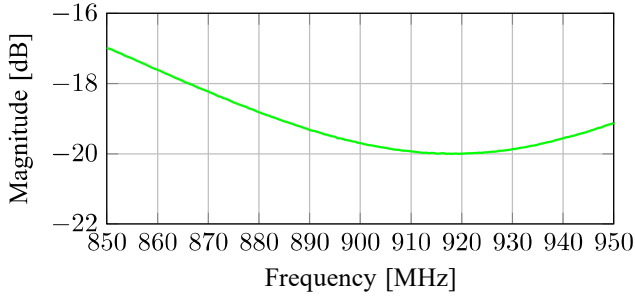

Fig. 3. S21 Circulator with antenna emulation.

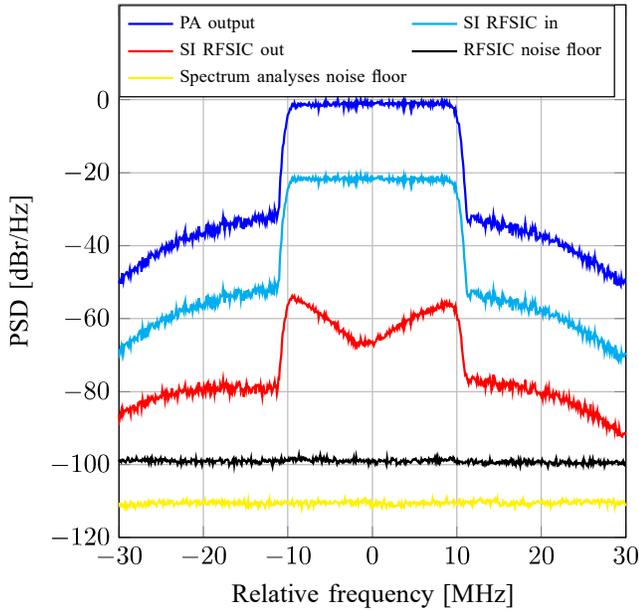

Fig. 4. SIC performance of the IBFD transceiver with RFSIC only.

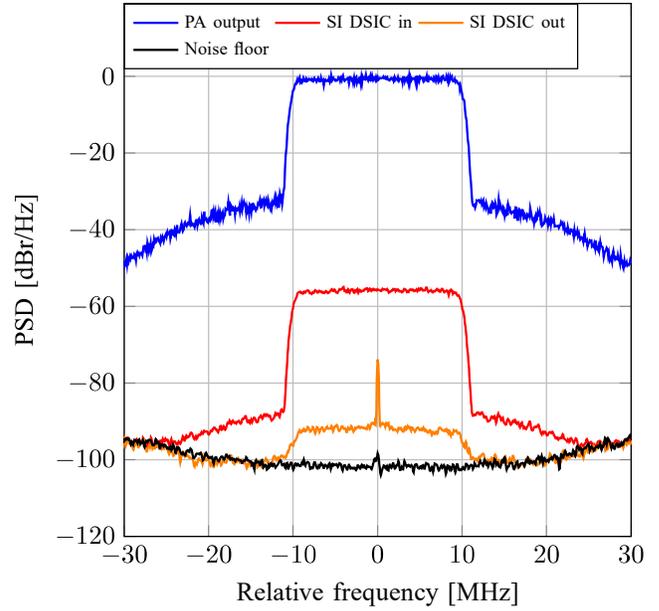

Fig. 5. SIC performance of the IBFD transceiver with DSIC only and an attenuated transmitted signal.

*2) DSIC performance:* This analysis was focused on the DSIC standalone case, where the RFSIC was switched off. In Figure 5, the SI signal over the different stages is shown. In order to prevent the saturation of the receiver front-end, the transmitted signal was attenuated properly as denoted by the DSIC input in the mentioned figure. In this test, the DSIC target is to cancel the linear and nonlinear components of the transmit signal. For the case after the DSIC, it is observed that not all interference can be canceled with the chosen configuration. Two reasons for this behavior are possible, either this remaining interference has a random source, or it is not covered by our signal model used for the cancellation. Due to the imperfections of the used receiver, the noise floor (in black) is not flat in the frequency domain.

*3) Combined performance with receive signal:* The final performance showcase is presented in Figure 6, where a SoI is received simultaneously while transmitting a signal. To illustrate the reception of the signal, we reduced the bandwidth of the signal transmitted from the signal generator to 10 MHz.

The SI is illustrated over the multiple cancellation stages. The RFSIC performance follows that of Figure 4, whereas after the DSIC is now dominated by the SoI. This proofs the system's capability to cancel the SI considerably. Such cancellation performance enables the success reception of the SoI regardless of the large power disparity between the transmitted signal and the SoI. A final summary of the achieved cancellation values is presented in Table II, where a total of 91 dB cancellation was realized.

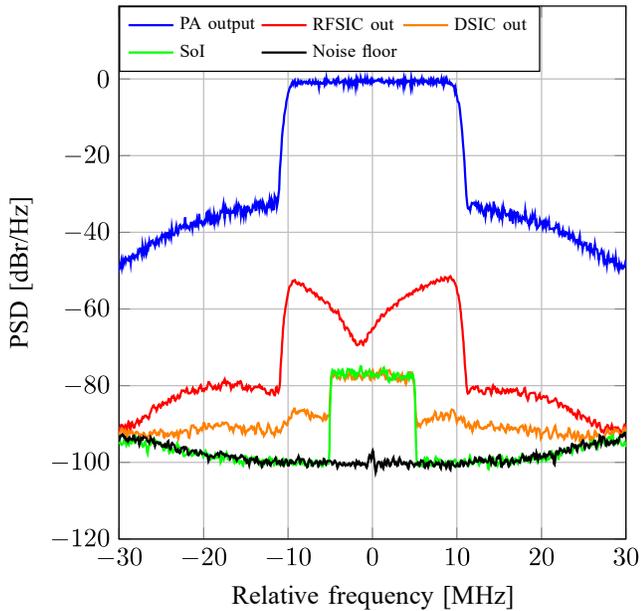

Fig. 6. SIC performance of the IBFD transceiver (RFSIC and DSIC) with received signal of interest.

TABLE II
CANCELLATION PERFORMANCE.

|  | Power [dBm] | Cancellation [dB] |
| --- | --- | --- |
| Transmit signal after PA | 20 | - |
| SI after circulator | -1 | 21 |
| SI after RFSIC | -37 | 36 |
| SI after DSIC | -71 | 34 |
| total cancellation | - | 91 |

## IV. CONCLUSION

This work presented an IBFD architecture based on a SDR-RF transceiver. The simple, digitally controlled RFSIC consists of a fixed delay and a vector modulator. A single antenna is used for transmission and reception in the investigated setup. The proposed solution for the DSIC is based on the exponentially weighted transversal RLS-DCD with a pre-orthogonalization stage of the nonlinear basis components. This solution was motivated by a good cancellation performance along with a low-complexity target for the digital canceler. The system's evaluation illustrates the effectiveness of the proposed cancellation solutions for both the RFSIC and the DSIC. The system setup was further extended to include a received SoI to emulate practical scenarios of full duplex systems deployment.

ACKNOWLEDGMENT

This work was supported by the French project DUPLEX and the European Commission in the framework of the H2020-ICT-2014-2 project Flex5Gware(Grant agreement no. 671563).